\documentclass{aastex63}

\usepackage{amsmath}
\accepted{April 10, 2022}
\submitjournal{AJ}
\usepackage{hyperref}
\hypersetup{
    colorlinks=true,
    linkcolor=blue,
    filecolor=magenta,      
    urlcolor=blue,
    pdftitle={Overleaf Example},
    pdfpagemode=FullScreen,
    }

\shorttitle{UVOT SNe Ia with FPCA Fitting}
\shortauthors{Devarakonda and Brown}

\graphicspath{{./}{figures/}}

\begin{document}

\title{Comparisons of Type Ia Supernova Light Curves in the UV and Optical \\ with the Swift Ultra-Violet/Optical Telescope}

\correspondingauthor{Yaswant Devarkonda}
\email{yaswantd@tamu.edu}

\author{Yaswant Devarakonda}
\affiliation{George P. and Cynthia Woods Mitchell Institute for Fundamental Physics $\&$ Astronomy, \\
Texas A$\&$M University, Department of
Physics and Astronomy, \\
4242 TAMU, College Station, TX 77843, USA}

\author{Peter J. Brown}
\affiliation{George P. and Cynthia Woods Mitchell Institute for Fundamental Physics $\&$ Astronomy, \\
Texas A$\&$M University, Department of
Physics and Astronomy, \\
4242 TAMU, College Station, TX 77843, USA}

\begin{abstract}

We examine the light curve parameters of 97 nearby Type Ia supernovae in the ultraviolet and optical using observations from the Swift Ultra-Violet/Optical Telescope. Our light curve models used a linear combinations of templates, which were based on Functional Principal Component Analysis of the optical photometry of SNe Ia. The weights for each principal component used in the fit capture certain aspects of the light curve properties. We find that there is little difference in the overall variability of principal component scores across filters. We also find correlations between certain filters and PC components, indicating that the UV and optical properties seem to be tied together. This is a promising step in UV SNe Ia analysis, and suggests that UV light curves may be used to infer optical properties, paving the way for future cosmological studies. 

\end{abstract}



\section{Introduction} \label{sec:intro}

Type Ia Supernovae (SNe Ia) are useful as cosmological tools due to their highly uniform optical luminosity, which correlates strongly with their optical light curve profile and color \citep{Pskovskii_1977,Phillips_1993,Riess_etal_1996_mlcs,Phillips_etal_1999}. This has led to their use in studies for distance measurements and constraints on cosmological parameters \citep{Riess_etal_1998,Perlmutter_etal_1999}.

In optical wavelengths, SNe Ia have a strong relationship between their peak luminosity and luminosity-decline rate, leading to their widespread usage as standardizable candles \citep{Branch_1998,Leibundgut_2001}. This process is generally done by calibrating the peak luminosity with distance-independent metrics such as the light curve shape. This can be done by measuring the decline rate for the first 15 days after maximum brightness ($\Delta m_{15}$), often in the B band \citep{Phillips_1993,Phillips_etal_1999,Garnavich_etal_2004}. Another common method is measuring the stretch of a light curve when fitting to a template \citep{Goldhaber_etal_2001}.

The luminosity-width relation shows that more luminous SNe Ia have broader light curves. The cause of this relation is believed to be the production of $^{56}Ni$ during the SNIa explosion, and the change in the temperature and ionization evolution as a result \citep{Nugent_etal_1995,Hoeflich_etal_1996,Mazzali_etal_2001,Kasen_Woosley_2007}.  $^{56}Ni$ drives much of the SNIa luminosity, and the light curve shape is greatly affected by opacity effects from$^{56}Ni$ and Fe-group elements \citep{Mazzali_etal_2007_Ia}.

Subsequent studies have looked beyond the optical to see if these strong correlations still hold true at other wavelengths. For example, the near-infrared (NIR) luminosity has been shown to be less susceptible to the effects of interstellar extinction while also having a diminished dependence on the shape of the light curve \citep{Krisciunas_etal_2000,Wood-Vasey_etal_2008,Mandel_etal_2011,Kattner_etal_2012}. Meanwhile, studies in the near-ultraviolet have shown that the correlations have much larger scatter \citep{Jha_etal_2006_U,Kessler_etal_2009, Brown_etal_2010,Milne_etal_2010,Brown_etal_2017}. Understanding the scatter at shorter wavelengths will be an important task, as it enables us to better predict SNe Ia properties at larger distances where their rest-frame UV emission will be redshifted into the optical. 

One issue with analyzing the UV light curves of SNe Ia is the lack of empirical templates at those wavelengths. In the optical, the most common optical fitting techniques are MLCS \citep{Riess_etal_1996_mlcs, Jha_etal_2007} and SALT3 \citep{kenworthy_etal_2021}. MLCS directly models the light curve data using vectorized templates over a hypothesized grid, whereas SALT3 uses spectral energy distributions to model light curves. Here, we use templates based on functional principal component analysis (FPCA) of optical SNe Ia light curves \citep{He_etal_2018}. We apply these templates to UV and optical observations of 97 SNe Ia to examine how well this technique can characterize light curves in a regime that it wasn't trained in, and if the fitted parameters can shed new light into the diversity of SNe Ia in the UV. 

In Section \ref{sec:methods} we describe our observations and methodology for template fitting. Then in Section \ref{sec:analysis} we apply statistical tests to compare the diversity and correlation of our fitted parameters in the UV and optical regimes. Finally, in Section \ref{sec:disc} we examine our results in detail and summarize our work in Section \ref{sec:con}. 

\section{Methods} \label{sec:methods}

\subsection{Observations} \label{ssec:observations}
Observations were done with the Ultra-Violet Optical Telescope (UVOT; \citealp{Roming_etal_2005})on the Neil Gehrels Swift Observatory \citep{Gehrels_etal_2004}. 
Observations were usually performed in all six medium-band UVOT filters: $UVW2$, $UVM2$, $UVW1$, $U$, $B$, and $V$.  See \citet{Poole_etal_2008} for calibration and filter curve details, and \citet{Brown_etal_2010} for the effective throughput for a SN Ia.  In total, 219 SNe Ia were observed by Swift UVOT between March 2005 and October 2016 (SN 2005am through 2016gxp, inclusive) with photometry available through the Swift Optical Ultra-Violet Supernova Archive (SOUSA; \citealp{Brown_etal_2014}). A 3" or 5" aperture was used to maximise the signal-to-noise and zeropoints were based on the updated calibration of \citet{Breeveld_etal_2011} and the time-sensitivity correction from July 2015.

Upon manual inspection of the galaxy-subtracted UVOT light curves, we selected 110 SNe Ia which had at least four data points in the $UVW1$ filter covering the peak and early decline. Due to the faintness of SNe Ia in the mid-UV, some did not have adequate light curves in $UVW2$ and $UVM2$, and some were saturated in the $U$, $B$, and $V$ filters at peak. The light curves were initially fit following the procedure in \citet{Brown_etal_2017}.  Template light curves were fit to our data, using the peak magnitude, time of maximum, and stretch (linear scaling in the time axis) as free parameters. We used the $B$ and $V$ template curves from MLCS2k2 \citep{Jha_etal_2007} and the $UVW2$, $UVM2$, and $UVW1$ light curves of SN~2011fe \citep{Brown_etal_2012_11fe}.  The $UVW1$ light curve of SN~2011fe was also used to fit the $U$ band light curves due to the similarity of the mean light curves of $U$ and $UVW1$ bands found by \citet{Milne_etal_2010} and the saturation of the Swift $U$ band light curve of SN~2011fe near peak.

\subsection{FPCA Model} \label{ssec:FPCA}

To model our sample, we used templates constructed from functional principal component analysis (FPCA, \citealp{He_etal_2018}), a statistical method that uses a linear combination of a mean function and a few principal component (PC) functions to represent the light curve. The coefficients associated with these functions are called PC scores. The mean light curve function and PC functions were trained using a data set of light curve photometry for 111 SNe Ia in the $BVRI$ bands. A four PC template was created for each band, as well as a generic (band-vague) template that combined the training data across bands that can be used to model light curves in any filter.

To retrain the FPCA model to include UV photometry would require more quality observations and an amount of work beyond the scope of this paper. Rather, we take the simpler step of using the band-vague template to fit the UV and optical light curves simultaneously. While the inclusion of the R and I band in the training set leads to an over-prediction of flux at later phases, we find (\ref{ssec:models}) that the quality of the fit is comparable between the UV and optical bands in the regions near peak flux. Using the same template to fit across all bands has the added benefit of each PC controlling the same aspects of each light curve, allowing for direct comparisons. In Section \ref{sec:disc} we examine possible future work with UV trained modeling.

\begin{figure}[ht]
    \centering
    \includegraphics[scale=0.5]{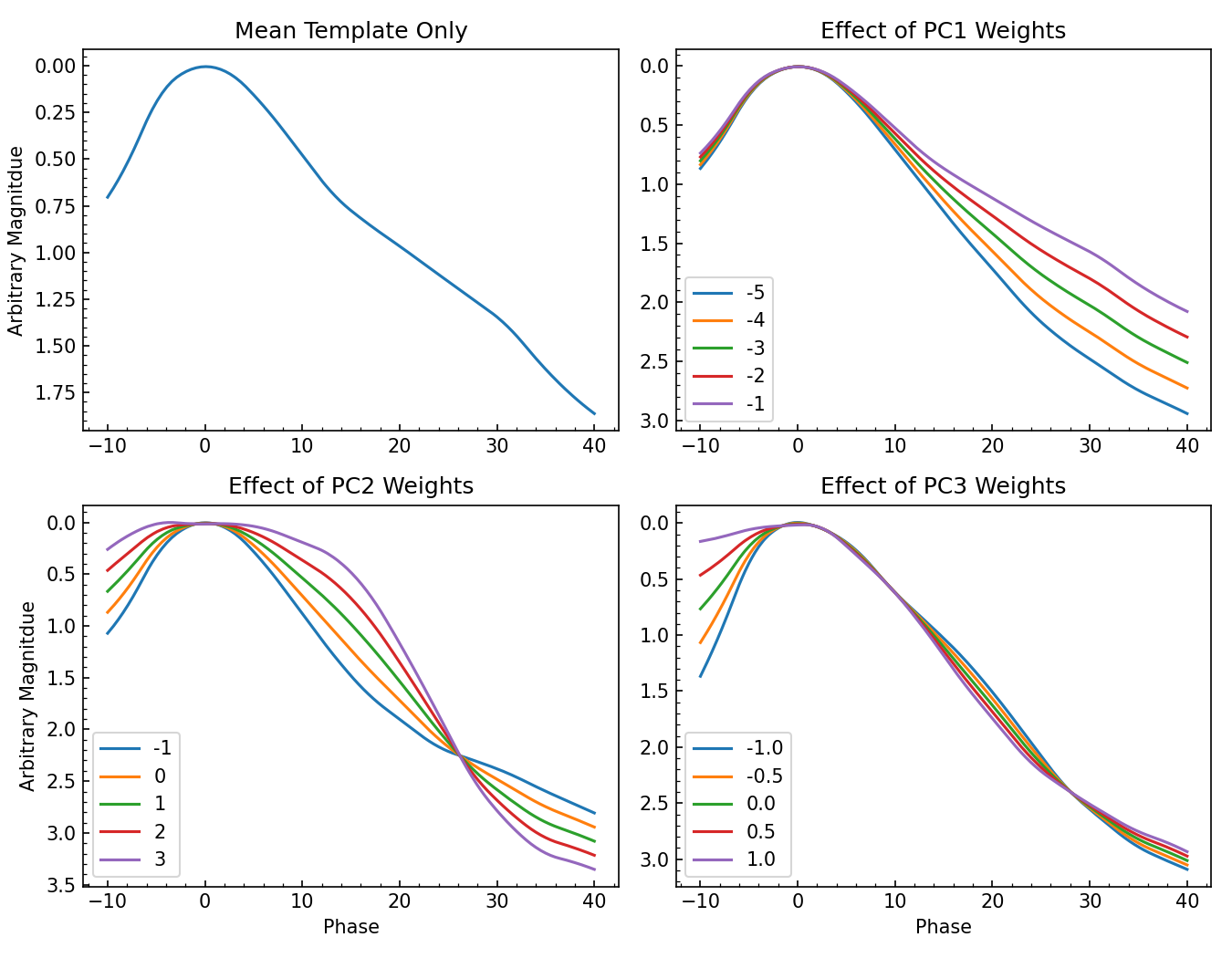}
    \caption{Examples of the effect that each PC template has on the light curve for a range of common B band PC score weights. The PC1 template mostly controls the late-phase decline rate, the PC2 template controls the overall stretch of the light curve and the post-peak decline rate, and the PC3 template has a plays a role in the pre-peak rise rate and the late phase flux.}
    \label{fig:score_comp}
\end{figure}

Per \citet{He_etal_2018}, the first PC reflects the decline rate after the peak, and the second PC reflects the light curve width around the peak. The third PC adjusts the pre-peak rise as well as a bump ~20 days after peak that is commonly seen in the near-IR bands, while the fourth PC accounts for more complex fluctuations. While the first three components are tied to certain light curve properties, they also describe a wider range of parameters. This means that while the first PC is related to the decline rate, it is not a perfect one-to-one correlation. In the original data set that this template was tested on the first PC score explained 92$\%$ of variability, the first two explained 96$\%$, and all four explained 99$\%$ of variability. In Figure \ref{fig:score_comp} we replicate the effects of a typical range of PC weights for visualization. 
\subsection{Template Fitting} \label{ssec:fitting}
We calculated the phase of each light curve in all six filters, corrected for redshift time dilation, by using the estimated peak from the preliminary modeling as our zero point.
\begin{equation}
    Phase_i = \frac{MJD_i-MJD_{peak}}{1+z}
\end{equation}

 We then removed data points that occur 10 days before the peak or 40 days after the peak, as those are outside of the domain of the template. We ensured that there was at least one data point before the peak and one at least 15 days after the peak, and that there was a minimum of five data points. The template models were then linearly interpolated to match the phases of the observations, and linearly combined to match the observed light curve with weights for the templates determined by the \texttt{curve\_fit} package from the SciPy library in python \citep{2020SciPy-NMeth}. The program attempts to fit the function to the data by minimizing the non-linear least squared errors. Below is an example of the function, 
 \begin{equation}
     LC(p) = M_{peak} + Mean_{template}(p) + \sum_{i=1}^N W_i * PC_i(p)
 \end{equation}

where $p$ is the phase, $M_{peak}$ is the peak magnitude (a free parameter, but given an initial guess based on the preliminary modeling estimates), $Mean_{template}$ is the mean light curve from the template, and $W_i$ is the weight of each principal component as determined by \texttt{curve\_fit}.

 \begin{figure}
     \centering
     \includegraphics[scale=0.5]{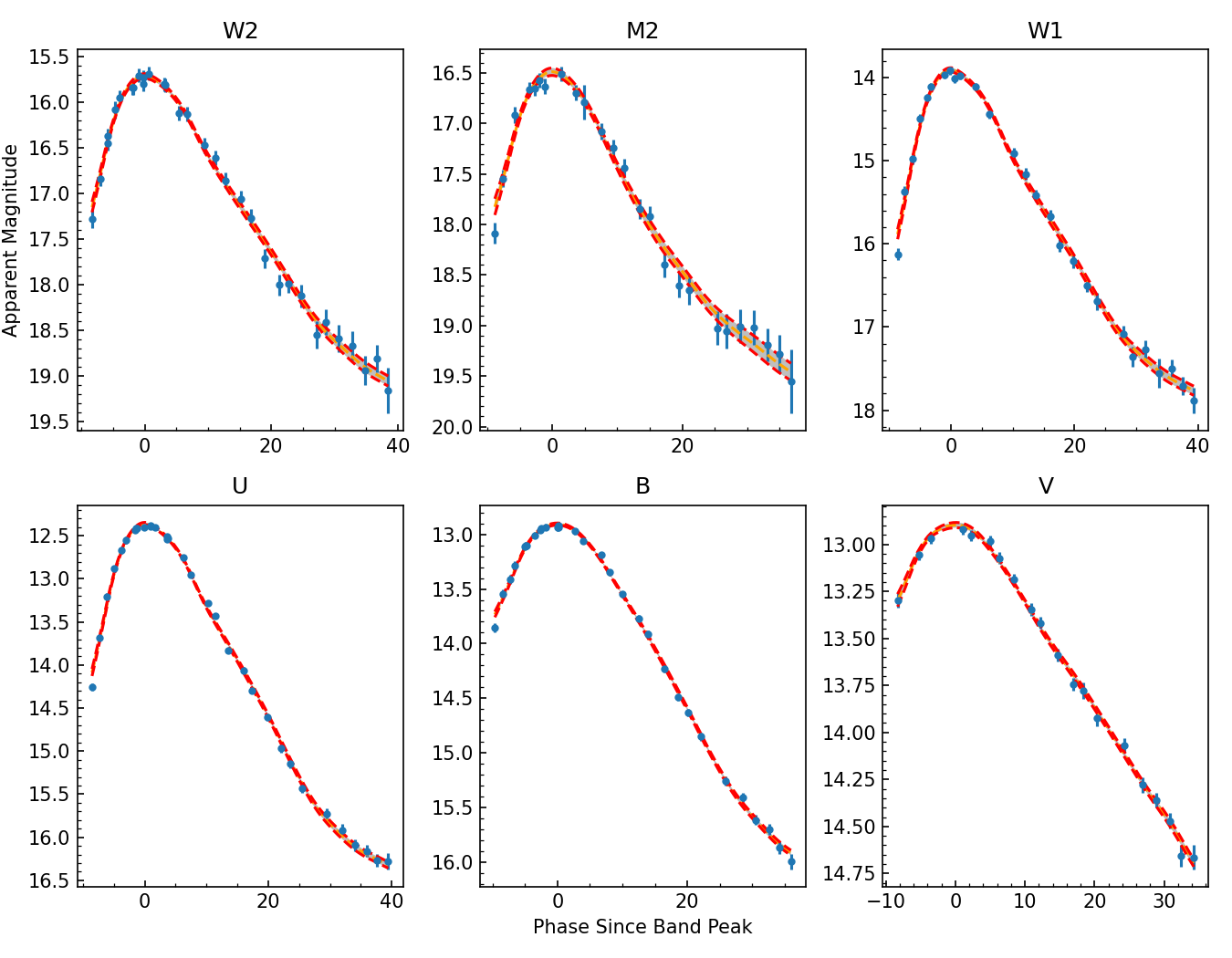}
     \caption{Swift observations of SN~2011by (blue) and the 3 template model (orange) with 1 $\sigma$ errors (red). A phase of 0 in each subplot corresponds to the date of the modeled peak brightness for each band.}
     \label{fig:by_initial}
 \end{figure}
 
\begin{figure}
    \centering
    \includegraphics[scale=0.85]{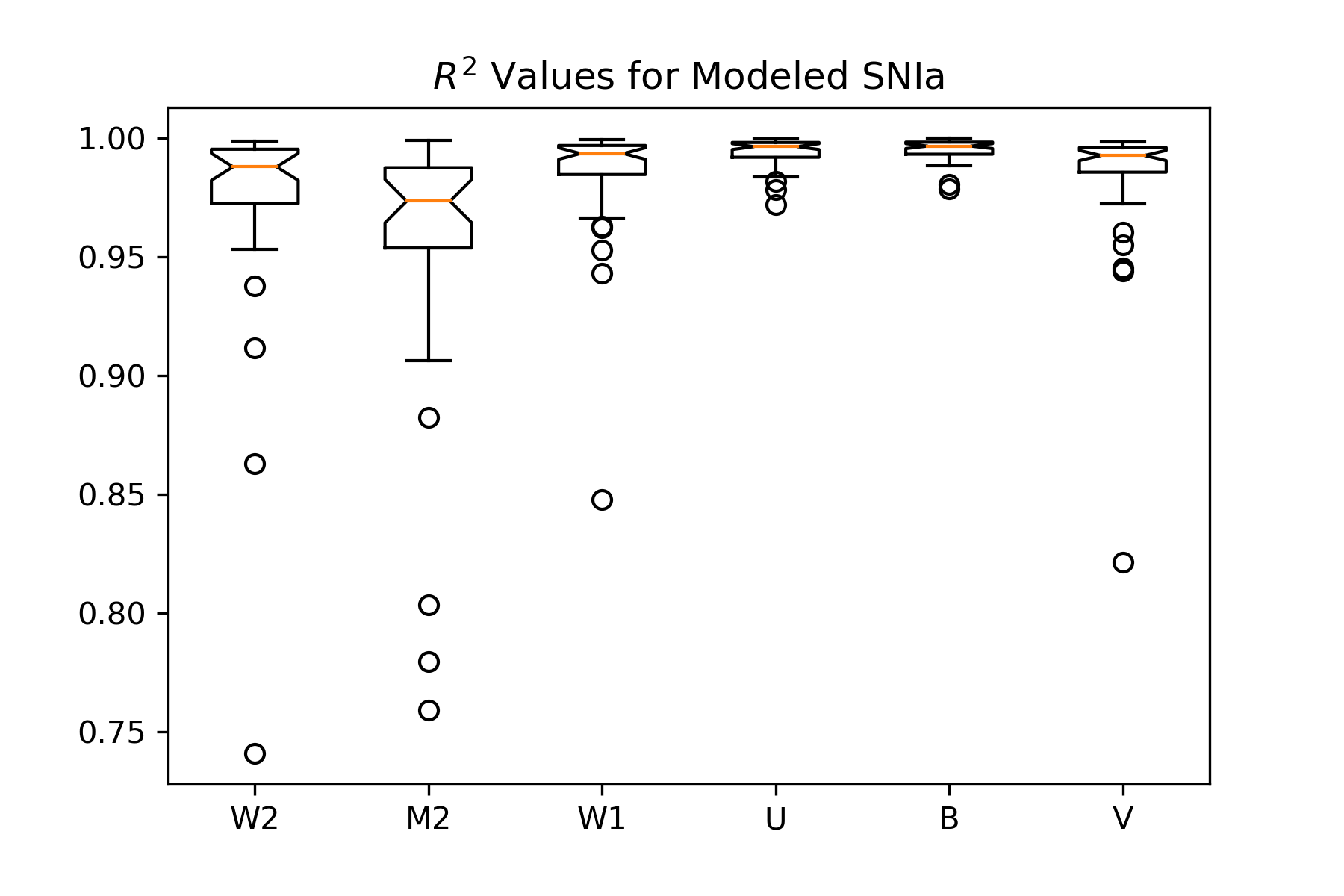}
    \caption{The $R^2$ values of modeled SNe Ia in each band. This value measures the amount of variance in the data that is explained by the model on a scale of 0-1. This measurement is done without accounting for the known observational error in the data, and is thus a lower limit on the true $R^2$ value. The notches on each box plot indicate the uncertainty in the median, and the whiskers indicate 1.5$\times$IQR distance.}
    \label{fig:r2_value}
\end{figure}

In total, there were 97 SNe Ia that had a fitted light curve in at least one filter, and 14 SNe Ia that had a fitted light curve in every filter. The $UVM2$ filter had the fewest well-fit light curves, due to the lower S/N of observations compared to other filters. Uncertainties in our model fits were determined by examining the full covariance matrix that results output by the \texttt{curve\_fit} package. We measured the $R^2$ values for the modeled SNe Ia (Figure \ref{fig:r2_value}) as a test of goodness-of-fit, and find that the majority of SNe Ia behavior is well explained by our model. The $U$ and $UVW1$ curves (which were not included in the training of the band-vague fitting) are fit with similar accuracy to the $B$ and $V$ band, respectively, and the PC fits explain 95\% of the points for 75\% of the mid-UV ($UVM2$ and $UVW2$) light curves.  

 \subsection{Model Selection} \label{ssec:models}
 To determine how many principal components were needed to ensure a good quality fit for our data set, we employed several tests of model selection. From the 14 SNe Ia with fits in all six bands, we selected four that had a relatively large number of observations (SN~2005ke, SN~2007af, SN~2007on, SN~2011by) to serve as test cases. Using the \texttt{train\_test\_split} package in the scikit-learn library in Python \citep{JMLR:v12:pedregosa11a}, we performed 100 bootstrap draws from each SNIa with a 70/30 split between the training and testing sample sizes. This allowed us to test the robustness of our fit with light curves that have sparse or inconsistent sampling, which is a more realistic case for the majority of our observations. 

\begin{figure}[ht]
\gridline{\fig{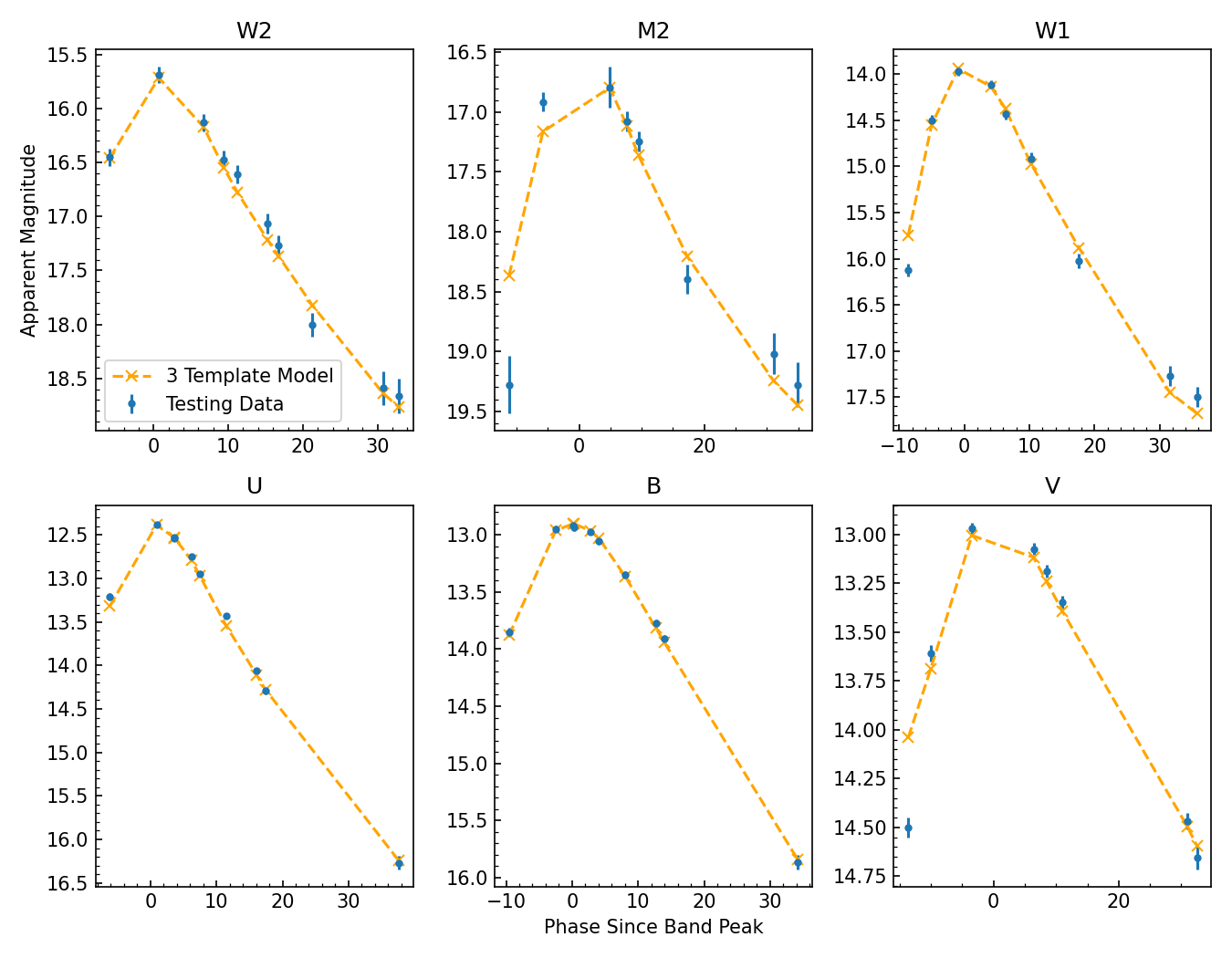}{0.5\textwidth}{Three PC Fit}
          \fig{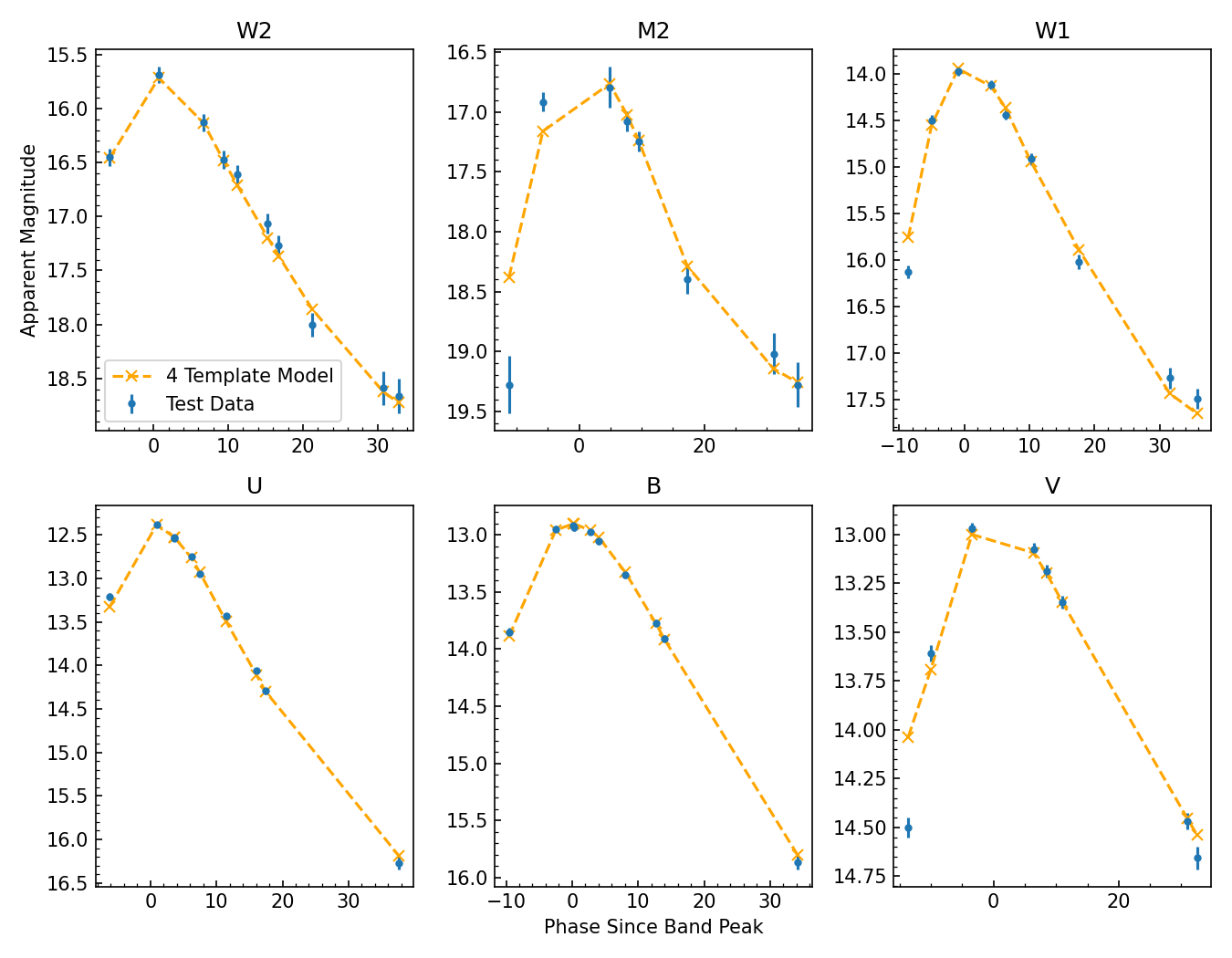}{0.5\textwidth}{Four PC Fit}
          }
\caption{An example of one of the bootstrap samples used for estimating the robustness of a three-PC model vs a four-PC model, using SN~2011by. After the model parameters are estimated using a training set, the model is then fit to a testing set and checked for goodness of fit. Increasing the number of templates used in the model from three to four does little to improve the fit of the model to the data.}
\label{fig:by_tests}
\end{figure}

 The Akaike Information Criterion (AIC) \citep{AIC} was calculated for each bootstrap sample and the mean AIC score for the two-, three-, and four-PC model were compared. The AIC score is a comparison test that measures how well a model fits a data set given a certain level of model complexity. In general, a lower AIC score indicates a better model. We employed a modified version of the AIC to account for the relatively small sample size, as shown in Equation 3. 

\begin{equation}
\begin{split}
     AIC &= 2k-2ln(\hat{L}) \\
     AIC_{small \ sample} &= AIC + \frac{2k^2 +2k}{n-k-1}    
\end{split}
\end{equation}
$\hat{L}$ is the maximum value of the likelihood function of the model, $k$ is the number of parameters, and $n$ is the sample size. We also explored using the Bayesian Information Criterion, but determined that our sample sizes were too small for an accurate assessment. 

Overall, we found that the two-PC and three-PC models had the lowest mean AIC scores in all six filters, signifying them as more appropriate models than the four PC model to use for our analysis. Between the two-PC and three-PC models, neither one had a significant difference when examining the larger data set. Future work with the FPCA template fitting could include individualized fitting for each SNIa and filter with more rigorous testing, but for uniform fitting we chose the three PC model. It added robustness to our modeling by improving our estimates of the rise of the light curves, with a minor risk of overfitting certain SNe Ia.

\begin{deluxetable}{ccccccc}[ht]
\tablenum{1}
\tablecaption{A comparison of the mean and standard deviation of the AIC scores taken from 100 random samples of SN~2011by data. In general, a lower AIC score indicates a more favorable model.  }
\tablewidth{0pt}
\tablehead{
\colhead{Filter} & \colhead{Training Set} & \colhead{Testing Set} & \colhead{3 Template} & \colhead{3 Template}
& \colhead{4 Template} & \colhead{4 Template} \\
\colhead{Type} & \colhead{Sample Size} & \colhead{Sample Size} & \colhead{Mean} & \colhead{$\sigma$} & \colhead{Mean} & \colhead{$\sigma$} 
}
\startdata
W2 & 22 & 10 & 5.4 & 0.7 & 12.3 & 0.7 \\
M2 & 18 & 8 & 6.3 & 1.4 & 17.0 & 1.8 \\
W1 & 18 & 9 & 4.9 & 1.3 & 12.8 & 1.4 \\
U  & 23 & 10 & 3.5 & 1.4 & 10.4 & 1.5 \\
B  & 23 & 10 & 3.0 & 1.7 & 9.6 & 1.9 \\
V  & 16 & 8 & 4.0 & 1.5 & 14.0 & 1.6 \\
\enddata
\end{deluxetable}

\section{Analysis and Results} \label{sec:analysis}
We employed a series of statistical tests to examine the similarity in the distribution of PC scores across the different bands. The Brown-Forsythe test \citep{brown-forsythe} examines the differences in the variance of two data sets by examining the distribution around the median. The null hypothesis for this test is that there is no statistical difference in the variance, while a p-value below 0.05 would indicate there is a difference in variance (Section \ref{sec:app}). 

\begin{figure}[ht]
    \centering
    \includegraphics[scale=0.57]{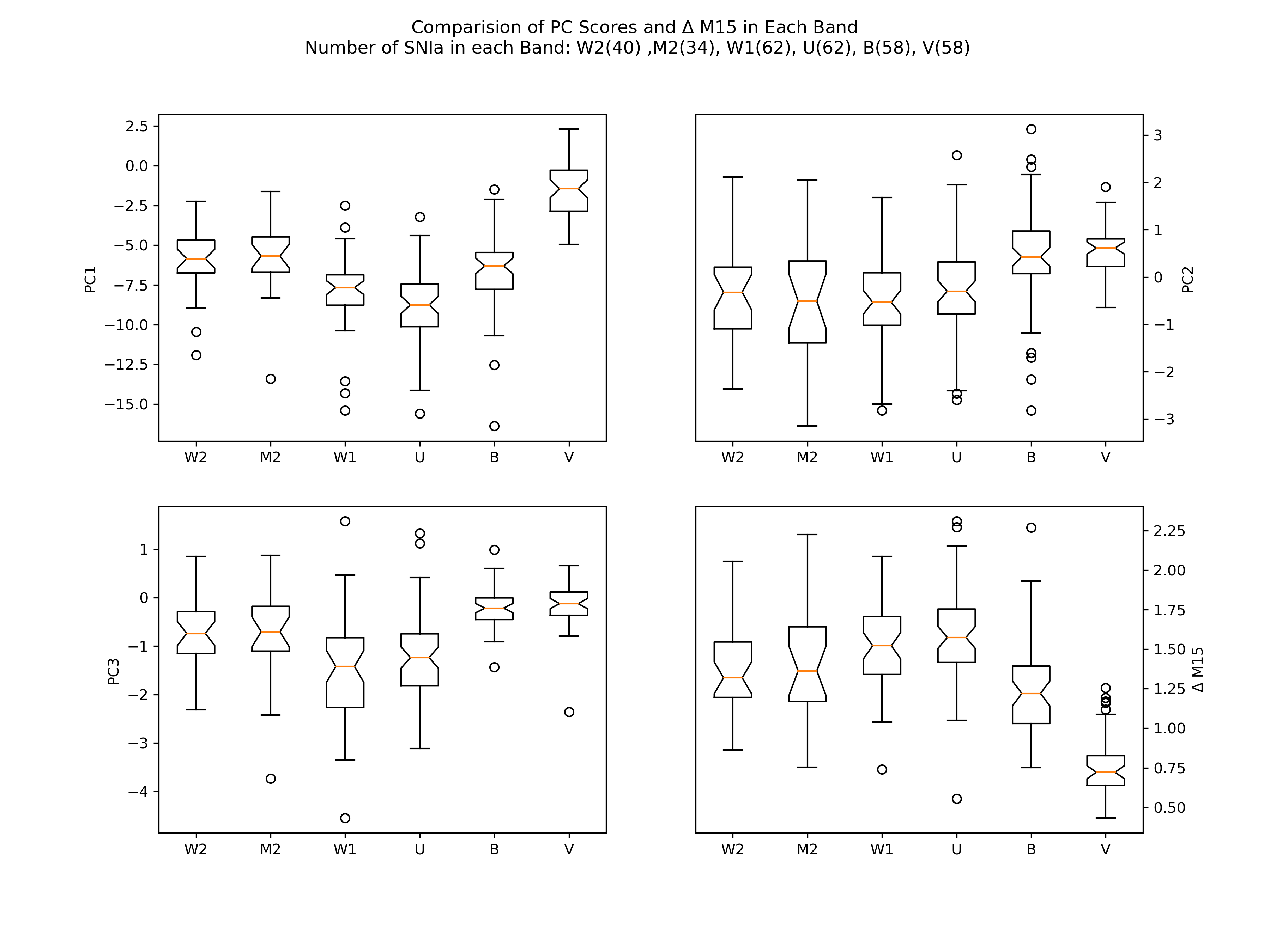}
    \caption{Box plots depicting the spread of parameters in each band. The notches indicate the uncertainty in the median, and the whiskers indicate 1.5$\times$IQR distance. Light curves with parameter estimates beyond the whiskers are labeled as circles.  }
    \label{fig:boxy}
\end{figure}

Additionally, we visually inspected the data variability with box plots (Fig. \ref{fig:boxy})\citep{tukey77}. Boxplots allow us to identify quickly and assess outliers, while also being less susceptible to issues such as sample size and skewness that may bias p-values. In particular, we examined the interquantile range (IQR) of the different PC scores in each band. Similar IQR scores would suggest that the populations have similar variances.  Visual analysis of the outliers revealed that many had models with acceptable fit quality, indicating that our model is capable of fitting the full diversity of our sample.

We find little statistical difference between the spread of PC1 and PC2 scores for both the UV and optical bands when compared to the $B$ band PC scores. For PC3 weights, only the $V$ band has a similar spread in values to the $B$ band, all other bands show a much larger spread in potential values. As both the PC1 and PC2 scores relate to the near-peak decline rate, nether can be solely used as a proxy for $\Delta m_{15}$. However the full model can be used to directly estimate the $\Delta m_{15}$, and the derived values show similar diversity with the $B$ band values (except for the $V$ band, which has a much tighter spread of values).

To examine the one-to-one correlations between filters, we plotted the parameter values in each filter versus the $B$ band value. We then estimated the Spearman rank-order coefficient \citep{Kokoska_Zwilinger_2000}, a non-parametric measure of two sets for a monotonic relationship. For parameters that have a p-value below 0.05, we test out whether a linear model can describe the relationship. While the true relationship may not be linear, having the linear fit can be a useful visual representation of the correlation and distribution of parameters. To perform this linear regression we employed the Python routine LINMIX, which uses Bayesian inference to return random draws from the posterior distribution \citep{Kelly_2007}. The routine uses a 1,000 iteration Markov Chain Monte Carlo procedure to converge on the posterior. The benefit of using this routine is its ability to handle uncertainties in both the independent and dependent variables as well as intrinsic scatter. From the iterations, we take the average value of the slope and intercept as our best fit parameters, and estimate the $r^2$ value to see how well the linear fit describes the correlation. The $r^2$ value is based on the total sum of squares ($SS_{tot}$) and the sum of squared residuals ($SS_{res}$) of the observed ($y$) and modeled ($f$) data, where

\begin{equation}
\begin{split}
    SS_{tot} &= \sum_i (y_i - \hat{y})^2 \\
    SS_{res} &= \sum_i (y_i - f_i)^2 \\
    r^2 &= 1- \frac{SS_{res}}{SS_{tot}}
\end{split}
\end{equation}

\begin{figure}[ht]
    \centering
    \includegraphics[scale=0.8]{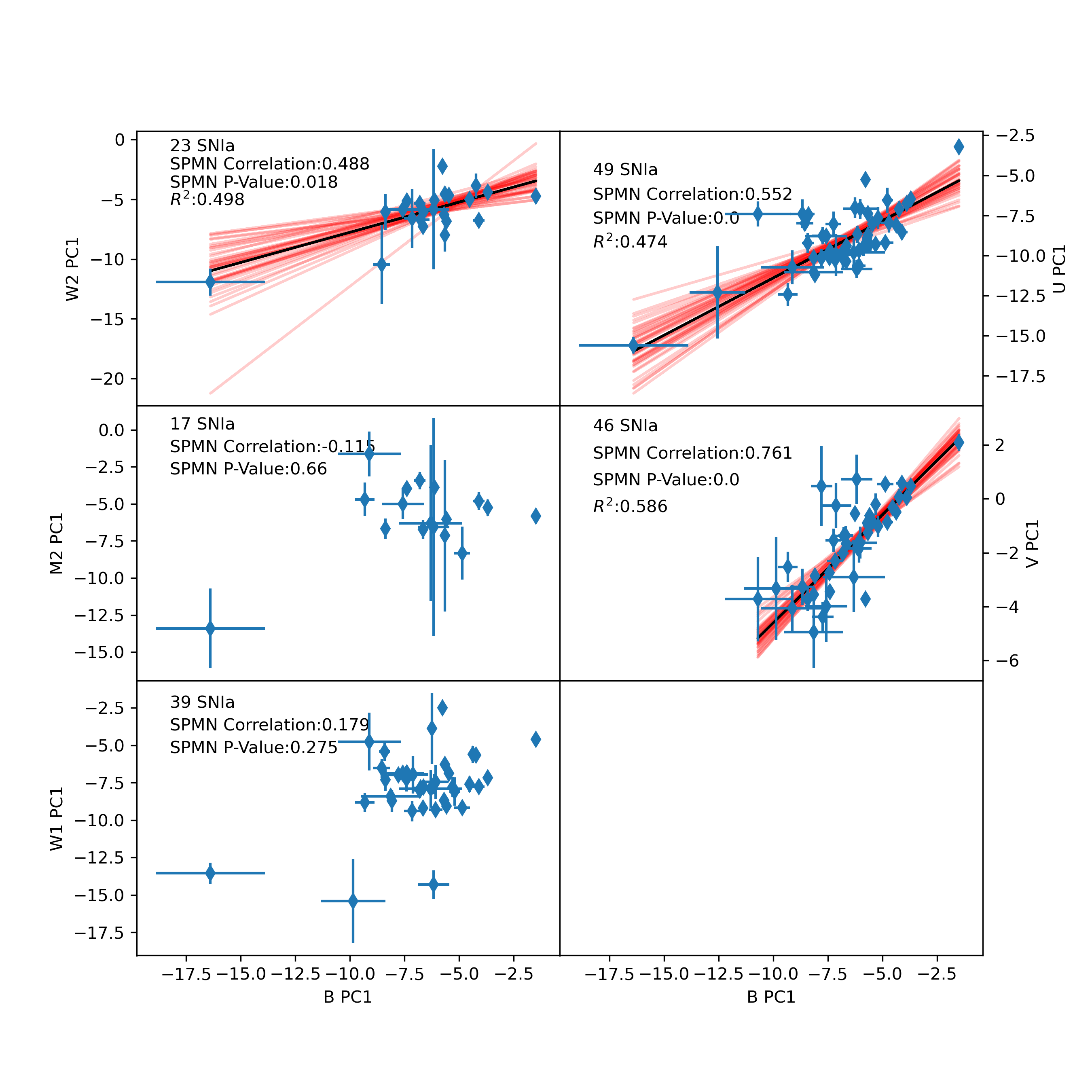}
    \caption{The correlation between the $B$ band and other Swift UVOT bands for the PC1 weights. Only the $UVW2$, $U$, and $V$ bands have a statistically significant correlation based on the Spearman rank-order coefficient test. The red lines indicate the spread of the MCMC draws from Linmix, and the black line is the mean linear fit from which we calulate the $r^2$ value. The PC1 weights are most directly related to the decline rate of the light curves.}
    \label{fig:pc1_stuff}
\end{figure}

\begin{figure}
    \centering
    \includegraphics[scale=0.8]{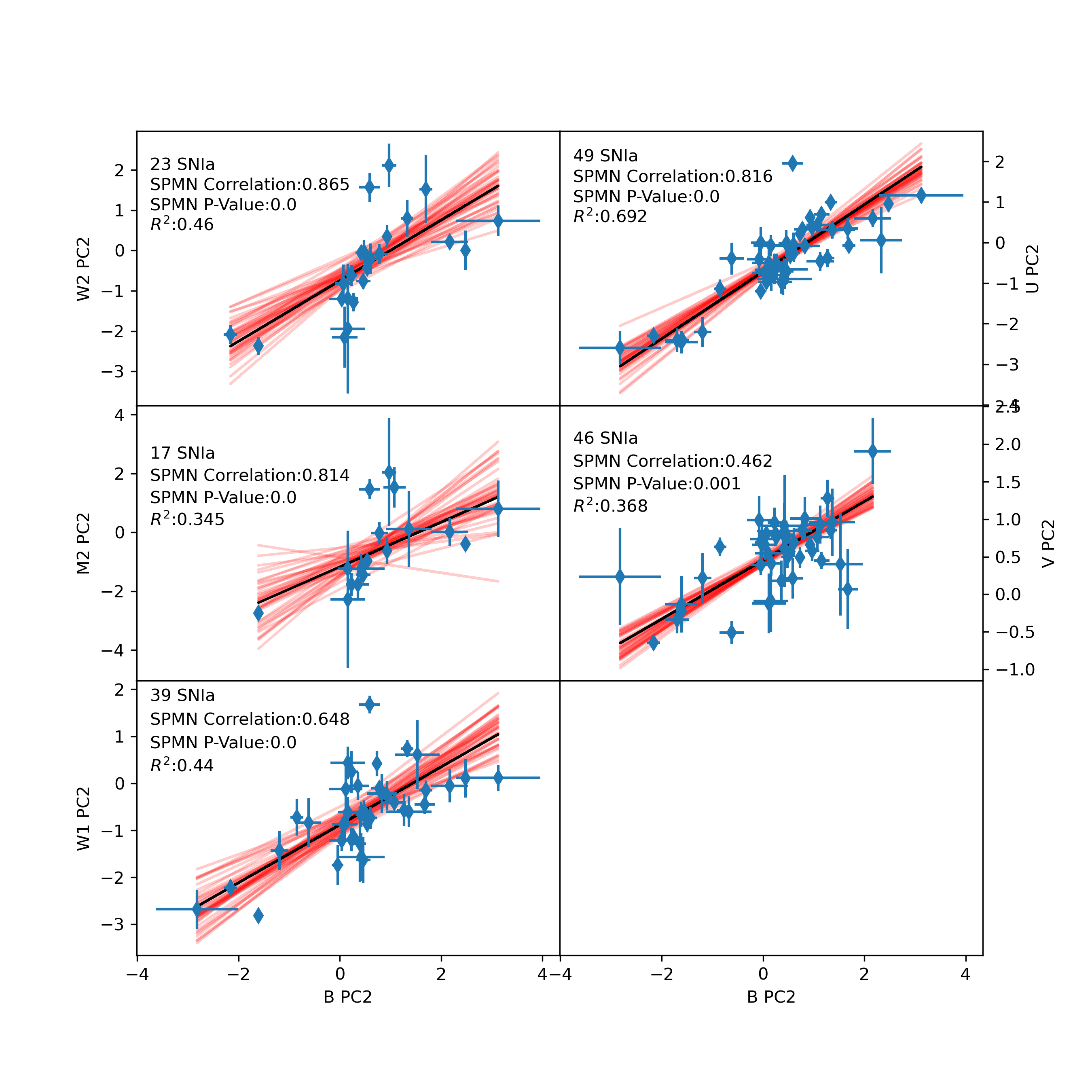}
    \caption{The correlation between the B band and the other Swift UVOT bands for the PC2 weights. The statistical significance of the correlation is based on the Spearman rank-order coefficient test. The red lines indicate the spread of the MCMC draws from Linmix, and the black line is the mean linear fit from which we calulate the $r^2$ value. The PC2 weights are most directly related to the spread of the light curves.}
    \label{fig:pc2_uv}
\end{figure}

We find that while the optical properties ($U$ and $V$ bands) are highly related to the $B$ band properties, only the PC2 weights (Fig. \ref{fig:pc2_uv}) show a consistent correlation between the mid-UV and near-UV bands and the $B$ band. For PC1 (Fig. \ref{fig:pc1_stuff}), there is a strong correlation between the $UVW2$ band and the $B$ band, which may be due to the relatively higher quality and quantity of data compared to the $UVM2$ and $UVW1$ bands.

\begin{figure}
    \centering
    \includegraphics[scale=0.8]{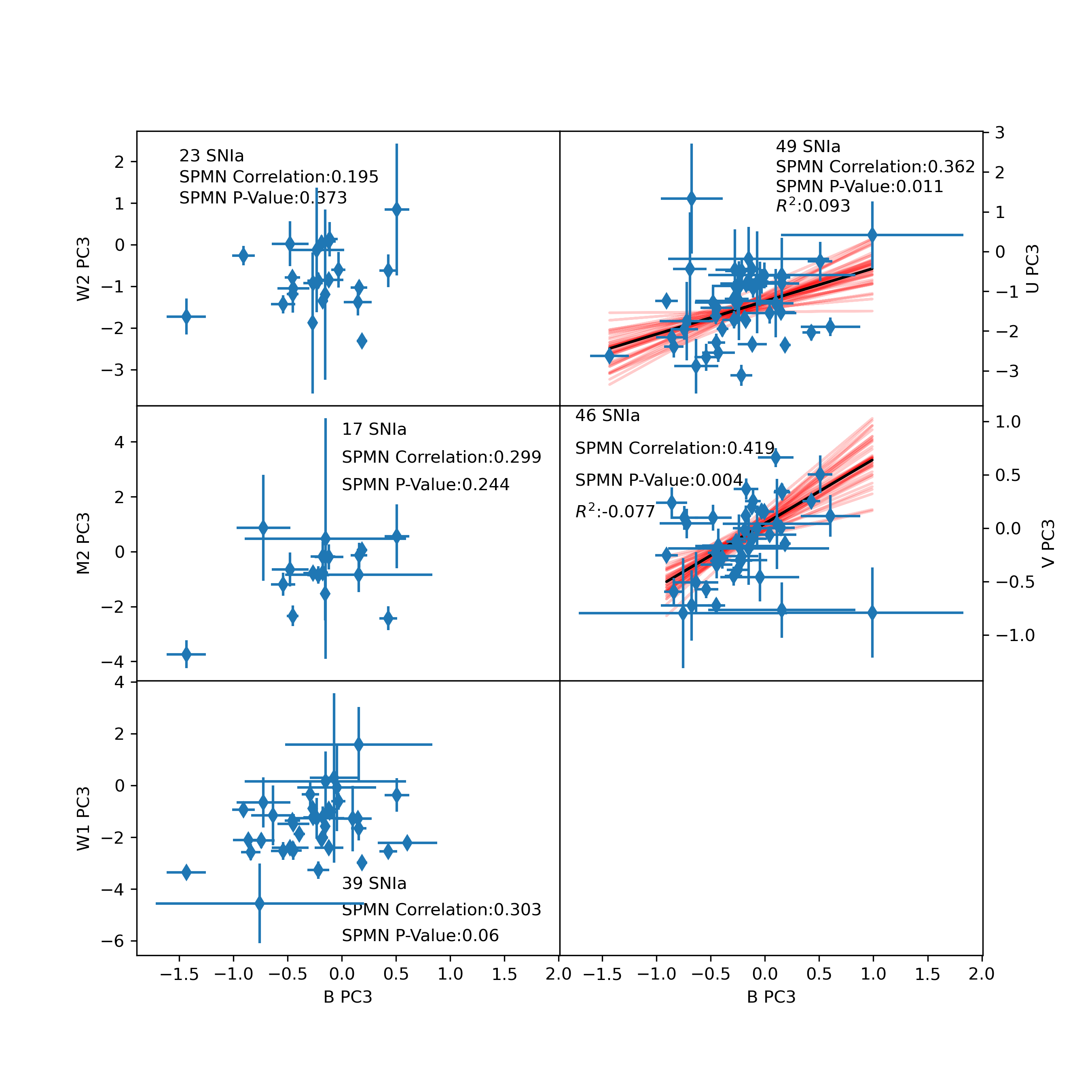}
    \caption{The correlation between the B band and the other Swift UVOT bands for the PC3 weights. Only the $U$ and $V$ bands have a statistically significant correlation based on the Spearman rank-order coefficient test. The red lines indicate the spread of the MCMC draws from Linmix, and the black line is the mean linear fit from which we calulate the $r^2$ value. The PC3 weights are most directly related to the pre-peak rise in the light curve and late phase adjustments.}
    \label{fig:pc3_stuff}
\end{figure}

\begin{figure}
    \centering
    \includegraphics[scale=0.8]{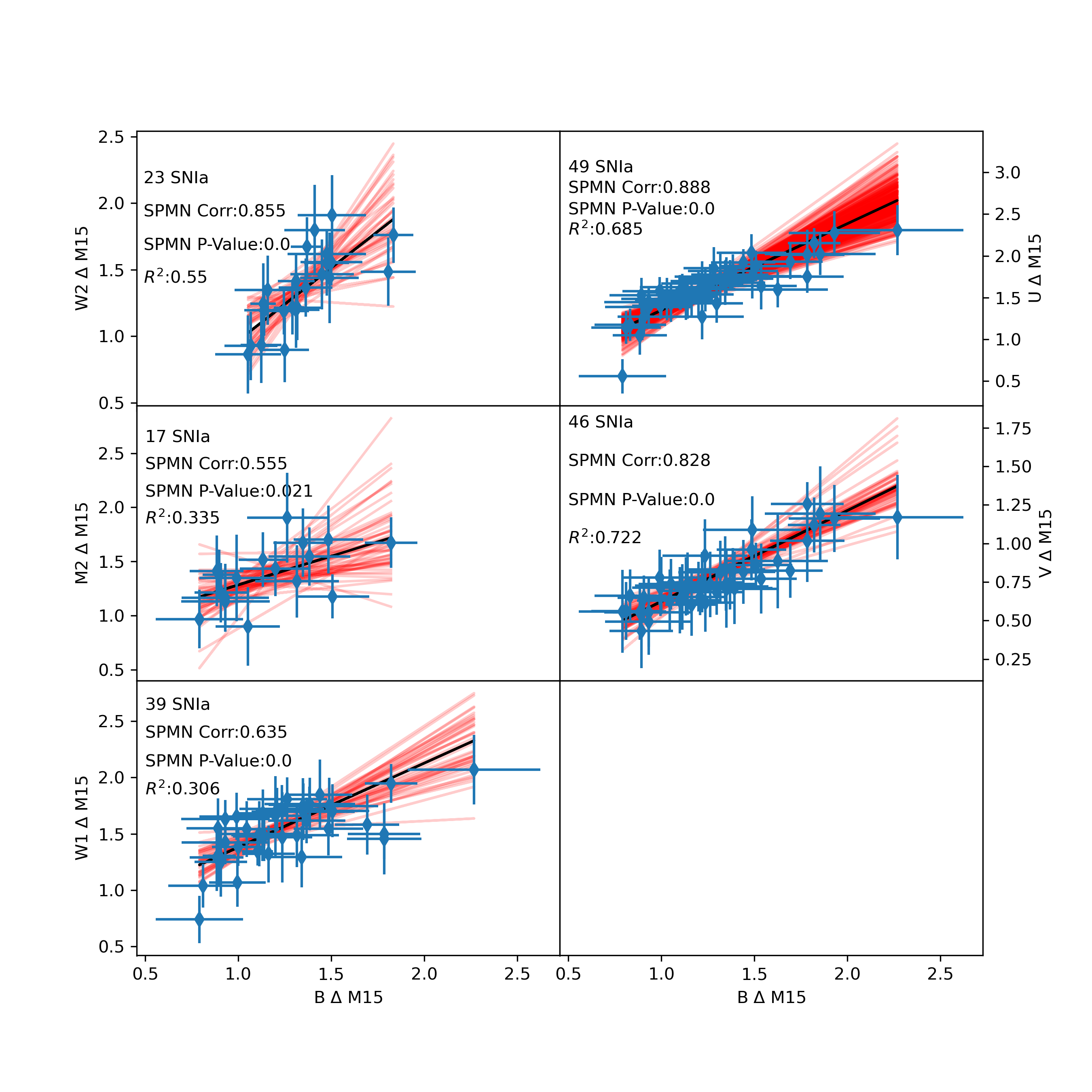}
    \caption{The correlation between the B band and the other Swift UVOT bands for the 15 day decline rate post-peak. The red lines indicate the spread of the MCMC draws from Linmix, and the black line is the mean linear fit from which we calulate the $r^2$ value. The decline rate is calculated from the best fit FPCA model for each individual light curve.}
    \label{fig:dm15_stuff}
\end{figure}

There is little to no statistical correlation for the PC3 weights (Fig. \ref{fig:pc3_stuff}) between the UV bands and the $B$ band, although the $UVW1$ could be a statistical edge case that lacks enough quality data to make a solid statement. The $\Delta m_{15}$ (Fig. \ref{fig:dm15_stuff}) shows a strong correlation and a relatively tight linear relation with the $B$ band in both the UV and the optical. 

\begin{figure}
    \centering
    \includegraphics[scale=0.7]{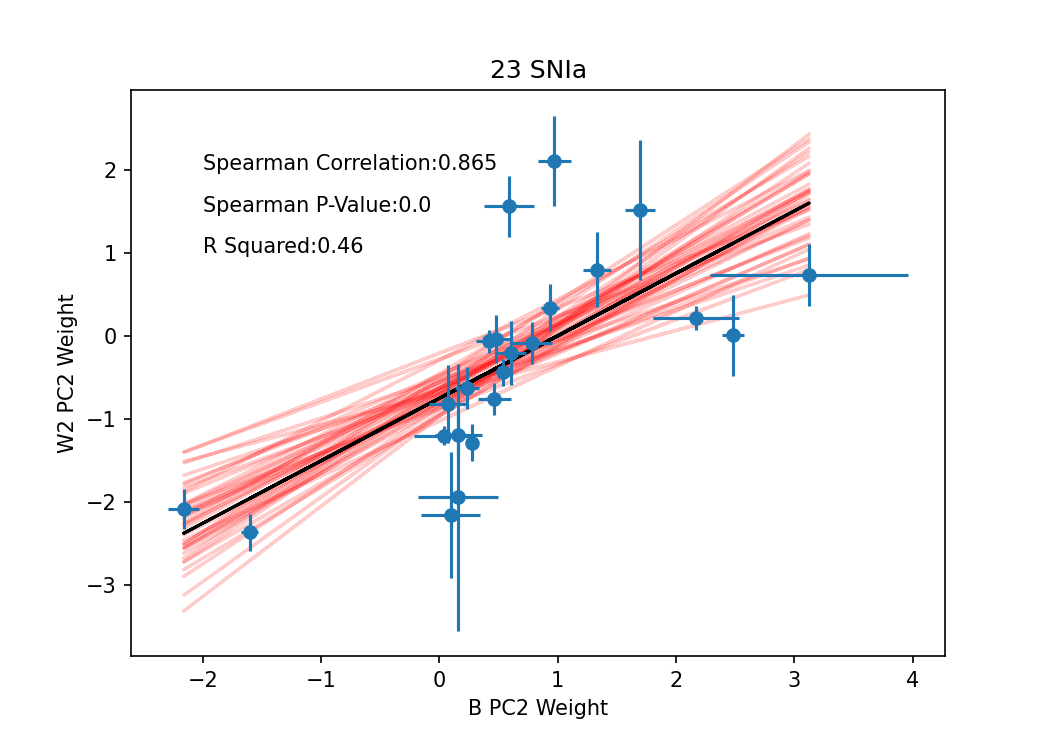}
    \caption{The comparison between the B band and W2 band PC2 weights for our sample of selected SNe Ia. The red lines indicate the range of possible regression fits from the LINMIX routine, and the black line indicates the linear fit based on the mean of the LINMIX coefficients. See Figures \ref{fig:pc1_stuff}, \ref{fig:pc2_uv}, \ref{fig:pc3_stuff}, and \ref{fig:dm15_stuff} in the appendix for the full set of correlated properties.}
    \label{fig:reggie}
\end{figure}

\begin{figure}
    \centering
    \includegraphics[scale=0.57]{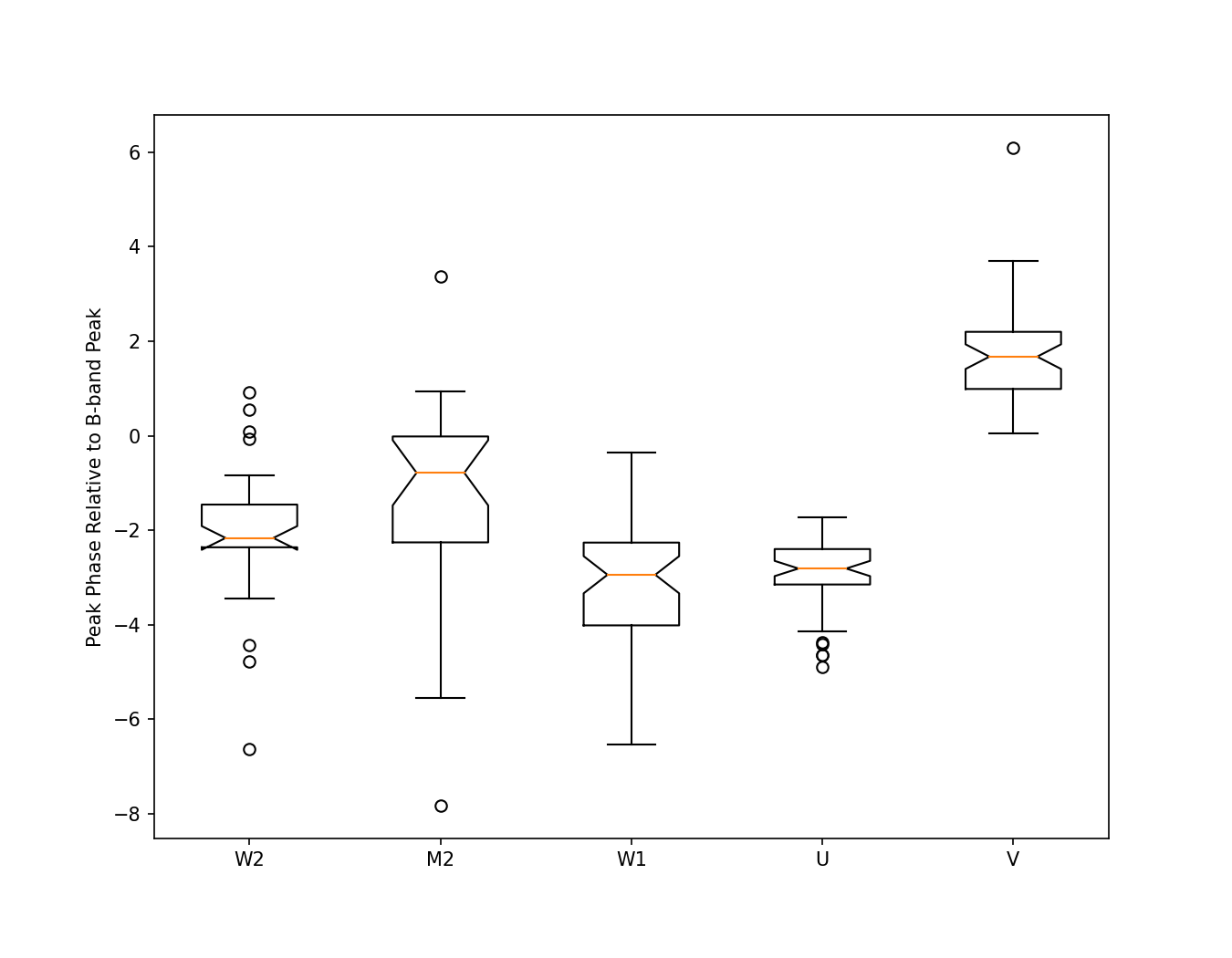}
    \caption{A boxplot of the peak phases for the light curves in each band relative to the $B$ band peak phase. Both the $UVW2$ and $U$ band light curves have a narrow distribution of peak phases at around 2 days before the B band peak phase, while the $V$ band light curves have a narrow distribution at around 2 days after the $B$ band peak phase. The $UVW1$ band light curves seem to peak earlier than the $B$ band light curves, but with a larger spread in the distribution.}
    \label{fig:peaks}
\end{figure}

Finally, we examined the differences in the peak magnitude phase between the $B$ band and the other filters, using a similar analysis to the PC score distributions. We find (Fig \ref{fig:peaks}) that the $U$ band and $UVW2$ band light curves have a tight spread of peaks phases at around 2 days before the $B$ band peak, and the $V$ band has a tight spread at around 2 days after the $B$ band peak (Figure \ref{fig:peaks}). The $UVM2$ and $UVW1$ band light curves also seem to peak earlier than the $B$ band light curves, although there is a much larger spread in the distribution of peak phases. 

Examining the outliers, we find that SNe Ia that peak in the UV 4 or more days before the B band peak were more likely to be classified as over-luminous in the literature. The most extreme outliers, SN~2011aa \citep{Kamiya_2012}, SN~2012dn \citep{Copin_etal_2012}, and LSQ12gdj \citep{Scalzo_etal_2014}, are all super-Chandrasekhar mass candidates. Thus, the difference in the time of peak in the UV and optical may be a useful metric for sub-type classification.

\section{Discussion} \label{sec:disc}

We successfully fit UV observations of SNe Ia using templates constructed from functional principal component analysis of optical photometry. While the quality of fit is better for the optical observations, quality of the UV fits were sufficient to perform a statistical analysis of the SNe Ia light curve properties. With improved UV observations and a larger data set, it should be possible to create filter-specific UV templates that will greatly improve future analysis of SNe Ia light curves. 

We find the the overall spread of the SNe Ia parameters is similar between the UV and optical bands in our sample, particularly for the PC2 template weights which are most directly tied to the stretch and decline rates of the SNe Ia light curve. In addition, we offer strong statistical evidence for a correlation between the UV and optical PC2 weights, suggesting that we can directly interpolate the spread of the SNe Ia light curve in one band based on the other. This will be highly useful with upcoming large survey telescopes such as the Rubin observatory, which is expected to find millions of SNe Ia at a wide range of redshifts. Interpreting the rest frame UV light from distant SNe Ia will require further understanding of the UV emission from nearby objects.

The lack of a strong correlation in the PC1 weights could suggest that physics behind the UV and optical light curves may be different at later phases ($\geq$20), where the PC1 weights have the largest impact. For phases immediately after the peak the PC2 weights have the strongest effect on the decline rate, resulting in the strong correlation with the $\Delta m_{15}$ rates. The correlations between PC2 weights in different bands suggests a common origin for the near-peak variations in the UV and optical light curves. Further studies with multi-epoch UV spectra of SNe Ia could provide more details into the emission mechanisms and the variations of SNe Ia light curves.

Previous work in the field has focused on rest-frame optical emission, leading to a wealth of robust models and large data sets. In comparison, there is a lack of quality observations and modeling in the UV. Our findings suggest that the UV and optical light curve properties may be more similar than previously suggested, which suggests the worth of further study into this regime. Future work will greatly constrain the true behavior of UV SNe Ia light curves relative to the better understood optical regime. 

These correlations will be important for the study of high redshift SNe Ia, whose rest-frame UV and optical emission will be simultaneously observable with upcoming ground-based telescopes such as the Vera Rubin Observatory. In the case that the optical light curve is poorly sampled (regardless of distance), the UV light curve may be able to provide constraints on optical properties such as the decline rate or stretch. The UV emission's potential for sub-type classification may, with further study, also assist in standardizing the optical relations. The UV has also been shown to significantly improve estimates on the amount of reddening due to dust extinction when used in conjunction with optical and NIR data \citep{Amanullah_etal_2015}, further showing the usefulness of multi-band studies.The diversity of the UV absolute magnitudes remains poorly constrained, however, hindering cosmological studies based only on UV observations.

\section{Conclusion} \label{sec:con}

We modeled the light curves of 97 SNe Ia in the UV and optical from the Ultra-Violet Optical Telescope on the Neil Gehrels Swift Observatory in order to better understand the correlation of their properties in different wavelength regimes. Modeling was done using a linear combination of band-vague templates, based on functional principal component analysis (FPCA) of a sample of SNe Ia in the optical bands. Each principal component is associated with a different light curve property, making it a useful tool for finding correlations between the filters.

We find that the overall dispersion in the PC1 and PC2 weights in all filters were similar, suggesting that the UV light curves are not as variable as the optical light curves. There is not a strong correlation in the PC1 weights for the UV and optical filters, which would indicate the lack of a strong correlation in the late-phase decline rate. There is a statistically strong correlation between the UV and optical filters for the PC2 weights, which would suggest that the stretch of the light curves and the post-peak decline rates are correlated across the wavelength regimes. This could point to underlying physics that ties the UV and optical flux together, with the bulk of the light curve shape likely being driven by $^{56}Ni$ decay. The differences that do exist are likely due to wavelength-dependent metallicity effects and the opacity of the SNe Ia atmosphere. This is the first study to statistically show that the UV and optical light curve properties have similar variability, and that the stretch and decline rate of the UV and optical light curves are significantly correlated. 

\section{Acknowledgements} \label{sec:ack}

The authors are grateful for the continuing support of the Mitchell Institute for Fundamental Physics and Astrophysics.
This work was supported by NASA grant 80NSSC20K0456, ``SOUSA's Sequel: Improving Standard Candles by Improving UV Calibration''.
The authors would like to thank Minjee Kim, Jianhua, Huang, and Xiaomeng Yan for lessons learned from related work. We also thank Nicholas Suntzeff, Lauren Aldoroty, and Jiawen Yang for their assistance and expertise, as well as the anonymous referee for their helpful comments. This work made extensive use of the Python Programming language and the following community-developed/maintained software packages: Jupyter \citep{jupyter}, NumPy \citep{numpy}, Matplotlib \citep{matplotlib}, Pandas \citep{pandas}, and SciPy \citep{Virtanen_etal_2020}. This
work also used \href{https://arxiv.org/}{arXiv.org} and NASA’s \href{https://ui.adsabs.harvard.edu/}{Astrophysics Data
System} for bibliographic information.


\bibliography{my}{}
\bibliographystyle{aasjournal}


\section{Appendix}\label{sec:app}

\begin{table}[ht]
\tablenum{2}
	\begin{center}
		\begin{tabular}{c|c|c|c}
			Band & Interquartile  Range & Brown-Forsythe Test Statistic & Brown-Forsythe P-Value \\
			\hline
			W2 & 2.062 & 0.169 & 0.682 \\
			M2 & 2.235 & 0.012 & 0.913 \\
			W1 & 1.913 & 0.112 & 0.738 \\
			U & 2.679 & 0.118 & 0.732 \\
			B & 2.320 & - & - \\
			V & 2.597 & 1.516 & 0.221 \\
		\end{tabular}
	\end{center}
	\caption{A comparison of the variability in the PC1 values as compared to the $B$ band. Similar IQR scores and P values $>$ 0.05 would suggest that there is not statistical difference in the variability between the bands.}
\end{table}

\begin{table}[ht]
\tablenum{3}
	\begin{center}
		\begin{tabular}{c|c|c|c}
			Band & Interquartile  Range & Brown-Forsythe Test Statistic & Brown-Forsythe P-Value \\
			\hline
			W2 & 1.304 & 0.426 & 0.516 \\
			M2 & 1.731 & 3.118 & 0.082 \\
			W1 & 1.110 & 0.359 & 0.550 \\
			U & 1.095 & 0 & 0.983 \\
			B & 0.900 & - & - \\
			V & 0.581 & 10.731 & 0.001 \\
		\end{tabular}
	\end{center}
	\caption{A comparison of the variability in the PC2 values as compared to the $B$ band. Similar IQR scores and P values $>$ 0.05 would suggest that there is not statistical difference in the variability between the bands.}
\end{table}

\begin{table}[ht]
\tablenum{4}
	\begin{center}
		\begin{tabular}{c|c|c|c}
			Band & Interquartile Range & Brown-Forsythe Test Statistic & Brown-Forsythe P-Value \\
			\hline
			W2 & 0.861 & 7.804 & 0.007 \\
			M2 & 0.925 & 13.237 & 0.001 \\
			W1 & 1.444 & 28.489 & 0 \\
			U & 1.075 & 21.533 & 0 \\
			B & 0.448 & - & - \\
			V & 0.481 & 0.062 & 0.804 \\
		\end{tabular}
	\end{center}
	\caption{A comparison of the variability in the PC3 values as compared to the $B$ band. Similar IQR scores and P values $>$ 0.05 would suggest that there is not statistical difference in the variability between the bands.}
\end{table}

\begin{table}[ht]
\tablenum{5}
	\begin{center}
		\begin{tabular}{c|c|c|c}
			Band & Interquartile  Range & Brown-Forsythe Test Statistic & Brown-Forsythe P-Value \\
			\hline
			W2 & 0.350 & 0.008 & 0.930 \\
			M2 & 0.473 & 0.107 & 0.745 \\
			W1 & 0.367 & 1.533 & 0.219 \\
			U & 0.337 & 0.031 & 0.861 \\
			B & 0.363 & - & - \\
			V & 0.188 & 9.904 & 0.002 \\
		\end{tabular}
	\end{center}
	\caption{A comparison of the variability in the $\Delta m_{15}$ values as compared to the $B$ band. Similar IQR scores and P values $>$ 0.05 would suggest that there is not statistical difference in the variability between the bands.}
\end{table}

\end{document}